\documentclass[aps,prd,twocolumn,nofootinbib,showpacs,superscriptaddress,groupedaddress]{revtex4-1}

\usepackage[T1]{fontenc} 

\usepackage{amsmath}
\usepackage{amssymb}
\usepackage{graphicx}
\usepackage{bm}
\usepackage{epsfig}
\usepackage{amsfonts}
\usepackage{color}
\usepackage{mathtools}
\usepackage[section]{placeins}

\newcommand{\be}{\begin{equation}}
\newcommand{\ee}{\end{equation}}

\newcommand{\bea}{\begin{eqnarray}}
\newcommand{\eea}{\end{eqnarray}}
\newcommand{\bean}{\begin{eqnarray*}}
\newcommand{\eean}{\end{eqnarray*}}

\newcommand{\xv}{{\mathbf x}}

\newcommand{\nn}{\nonumber}

\begin{document}

\title {Nonrelativistic lattice study of stoponium}
 
\author{Seyong Kim}
\affiliation{Department of Physics, Sejong University, Seoul 143-747,
  Korea}
\affiliation{School of Physics, Korea Institute for Advanced Study, Seoul 130-722, Korea} 

\begin{abstract}
We calculate the bound state properties of stoponium using lattice
formulation of nonrelativistic effective field theory for stop
which is moving nonrelativistically in the rest frame of
stoponium. Our calculation method is similar to that employed in
lattice nonrelativistic quantum chromodynamics (NRQCD) studies for
charmonium and bottomonium. Using $16^3 \times 256$ quenched lattice
gauge field configurations at $a^{-1} = 50(1) {\rm GeV}$, we obtain
the stoponium mass and the lattice matrix element which is related 
to the wavefunction at the origin for the $1S$ state and find that the
lattice $|R_{1S} (0)|^2/M_{1S}^3$ is $3.5 \sim 4$ larger than that
from a potential model calculation for $200 {\rm GeV} \le M_{1S} \le
800 {\rm GeV}$.
\end{abstract}

\maketitle


\flushbottom


\section{Introduction}
\label{sec:intro}

After the discovery of Higgs particle by ATLAS \cite{Aad:2012tfa} and
CMS \cite{Chatrchyan:2012xdj} at LHC, detailed measurements of its
property become very important and urgent. These precision
measurements will open up a new opportunity to search for physics
beyond Standard Model. In particular, heavy particles which can decay
into Higgs boson are under active experimental and theoretical
investigations. Scalar top quark (stop), the supersymmetric partner of
top quark and the Next-to-Lightest Supersymmetric Particle (NLSP), is
one of such possibilities. Then, stoponia, bound states (binding
through $SU(3)$ color gauge interaction) of stop and anti--stop may
become interesting provided that stop is long-lived enough to form a
bound state. They can serve as a probe to stop, and decay of stoponium
via the stop--anti stop pair annihilation into diboson states may be
observed due to its distinct signature \cite{Drees:1993yr,
  Drees:1993uw, Martin:2008sv, Kumar:2014bca, Batell:2015zla}.

With regard to the production cross section calculation of stoponium,
a next-to-leading order computation of perturbative part of the
production cross section is available \cite{Younkin:2009zn,
  Martin:2009dj} and a resummed next-to-next-to-leading logarithm
calculation is performed \cite{Kim:2014yaa}. Turning these
perturbative calculations into phenomenological comparisons requires
matrix elements for stoponium. With the observation that heavy quark
moves slowly in the rest frame of quarkonium ($v^2 \sim 0.1$ in
bottomonium and $v^3 \sim 0.3$ in charmonium where $v$ is the heavy
quark velocity in the rest frame of quarkonium), nonrelativistic
quantum chromodynamics (NRQCD) is developed and the quarkonium
production cross section is given in terms of perturbative parts and
nonperturbative matrix elements \cite{Bodwin:1994jh}. Similarly, the
parton-level differential cross section for the stoponium ($\Psi$)
production in a collider can be given as
\begin{equation}
d\hat{\sigma} (a b \rightarrow \Psi + X) = \sum_n d \hat{\sigma} (a b
\rightarrow \tilde{t} \overline{\tilde{t}} [ n ] + X ) \langle {\cal
  O}^\Psi [n] \rangle ,
\label{production_rate}
\end{equation}
where $a, b$ are partons, $\tilde{t}, \overline{\tilde{t}}$ are stop
and anti stop, $n$ denotes the angular momentum of the stoponium
states and ${\cal O}^\Psi [n]$ are generic forms of nonrelativistic
stoponium production operators \cite{Younkin:2009zn} by considering
nonrelativistic effective field theory (NREFT) for stoponium system
since in the rest frame of stoponium for $M \sim {\cal O} (100)$ GeV,
$v^2$ is expected to be $\sim {\cal O} (0.01)$ and $\alpha_s (Mv) \sim
{\cal O} (0.1)$.

Following \cite{Bodwin:1994jh}, one can relate the production matrix
elements to decay matrix elements by crossing relation in leading
order of $v^2$. The decay matrix elements then can be related to
nonrelativistic wavefunctions in the vacuum saturation approximation
upto ${\cal O} (v^4)$. For the $1S$ state stoponium, 
\begin{equation}
\langle {\cal O}^\Psi \rangle \simeq \langle 0 | \chi^\dagger \psi |
1S \rangle \langle  1S | \psi^\dagger \chi | 0 \rangle = |\langle 0 |
\chi^\dagger \psi | 1S \rangle |^2 , 
\end{equation}
when the leading $v^2$-order term in the factorization expansion is
considered and a stoponium state is assumed to dominate in the
intermediate states \cite{Bodwin:1994jh}. Here, $\psi$ denotes
nonrelativistic stop, $\chi^\dagger$ denotes nonrelativistic anti
stop. This matrix element is related to the wavefunction at the origin
\cite{Bodwin:1994jh} as
\begin{equation}
|\langle 0 | \chi^\dagger \psi | 1S \rangle |^2 \simeq
\frac{4\pi}{N_c}|R_{1S} (0)|^2  
\label{matrixwaveft}
\end{equation}
and has been calculated on lattice for charmonium and bottomonium
decays\footnote{Note that in Eq. \eqref{matrixwaveft} the factor
  $\frac{1}{N_c}$ instead of $\frac{1}{2 N_c}$, due to the spin-less
  nature of stop} \cite{Thacker:1990bm, Bodwin:1996tg,
  Bodwin:2001mk} (see \cite{Colquhoun:2014ica} for an improved lattice
calculation for bottomonium system).

So far, potential model estimates (e.g., \cite{Hagiwara:1990sq}) have
been used for the stoponium masses and the stoponium wavefunctions
where typical bound states properties are summarized in $\frac{|R_S
  (0)|^2}{{M_S}^3}$ with the mass of the S-wave stoponium state,
$M_S$, and $|R_S (0)|$ the value of stoponium radial wavefunction at
the origin. In $\frac{|R_S (0)|^2}{{M_S}^3}$, uncertainty from the
wavefunction at the origin, $|R_S (0)|$, is more dominant than that
from the mass. The mass of stoponium will be mostly from twice of the
``free'' stop mass and the ``binding energy'' will be just a few
percent in stoponium mass and the uncertainty in binding energy will
lead to sub-percent level uncertainty in stoponium mass. On the other
hand, $|R_S (0)|$ depends very much on the functional form of
potential models \cite{Eichten:1979ms, Kwong:1987mj, Thacker:1990bm}.

Clearly, using potential model estimates for the binding interaction
of stoponium is unsatisfactory since it introduces model-dependence
which can not be systematically improved, and makes the stoponium
cross section calculation unsystematic as a whole, despite the
improved perturbative calculations. Furthermore, potential models have
a difficulty in obtaining right decay widths for a given state
although they do better for relative ratios for different states
\cite{Eichten:1979ms, Kwong:1987mj}. Employing different functional
forms for potential models can lead to a large differences. For
example, a potential model estimate for $|R_S(0)|$ defies a naive
expectation that large stop mass ($M > 100$ GeV) would result in
Coulombic behavior of the wavefunction, and exhibits substantial
departure from the Coulombic value of $|R_S (0)|$ even at $M \simeq 1$
TeV \cite{Hagiwara:1990sq}. This suggests that there can be sizeable
non-perturbative contribution to the bound state properties. Indeed,
next-to-next-to-next-to-leading order calculation together with a scheme
choice which is less sensitive to the long distance effect of QCD is
necessary to understand the threshold behavior of the top anti-top
S-wave pair production cross section \cite{Beneke:2015kwa}.

In this work, we use lattice formulation of $v^2$ NREFT for stoponium
and calculate the stoponium mass and a stoponium matrix element,
$|\langle 0 | \chi^\dagger \psi | \Psi \rangle |^2$. Unlike a
potential model calculation, lattice NREFT is based on the first
principles of quantum field theory and allows systematic study of
errors associated with a given result. This effective lattice theory
is similar to the lattice version of NRQCD \cite{Thacker:1990bm,
  Lepage:1992tx} which allows highly successful understanding of
nonperturbative quarkonium physics (see e.g.,
\cite{Colquhoun:2014ica}) except for the fact that stop is a spin-less
particle. Our calculation for the stoponium property is performed on
$N_s^3 \times N_\tau = 16^3 \times 256$ lattices generated with
``quenched approximation'' and the lattice spacing, $a^{-1} \simeq 50$
GeV for the stop mass range $1 \le Ma \le 8$ (i.e, $50 \le M \le 400$
GeV). A large $N_\tau$ is necessary to avoid deconfining effects. In
the rest frame of stoponium, stop is expected to move slowly with
velocity $v ( \ll 1)$ and the size of stoponium should be smaller than
those of typical quarkonia (rough estimate for the size of stoponium
may be given by the self-consistency relation, $v \sim \alpha_s (M v)$
with the size of $r \sim (Mv)^{-1}$ and the stop mass $M$) and thus
small $N_s$ does not cause a significant finite volume effect. Since
the momentum scale larger than the heavy particle mass is ``integrated
out'' in NREFT and $M a$ is chosen to be $\sim 1$ , we need to
consider only the lattice spacing scale in lattice NREFT for
stoponium.

We find that with ${\cal O} (v^2)$ NREFT Lagrangian, the lattice
result for $|R_{1S} (0)|^2/M_{1S}^3$ is factor $3.5 \sim 4$ larger
than a potential model estimate in \cite{Hagiwara:1990sq} for $200
{\rm GeV} \le M_{1S} \le 800 {\rm GeV}$. Although further study is
necessary (in particular in view of the difficulty associated with
quantifying the systematic uncertainty from the quenched
approximation), this implies that the stoponium production rate at LHC
may be larger than the current estimates based on potential model. In
the following, we briefly summarize the lattice method used in the
calculation of stoponium properties (section \ref{sec:method}). Then,
we present our main result in Sec. \ref{sec:result} and conclude with
Sec. \ref{sec:summary}.

\section{Method}
\label{sec:method}

The effective Lagrangian for nonrelativistic stop in leading order of
$v^2$ is given by
\begin{equation}
\label{LNRQCD}
{\cal L} = \psi^\dagger \left(D_\tau - \frac{{\mathbf D^2}}{2M} \right) \psi +
\chi^\dagger \left(D_\tau + \frac{{\mathbf D^2}}{2M} \right) \chi,
\end{equation}
where $\psi$ is a complex scalar field for stop which transforms as
a $SU(3)$ vector and $\chi$ is that for anti-stop and $D_\tau$
(${\mathbf D}$) are gauge covariant temporal (spatial) derivative
under the strong interaction, $SU(3)$. Note that this leading order
NREFT Lagrangian does not differ from that for NRQCD. The difference
between stoponium NREFT Lagrangian and NRQCD Lagrangian occurs only
when next-to-leading order NREFT Lagrangian is considered since the
spin interaction (e.g., ${\mathbf \sigma} \cdot {\mathbf B}$) in NRQCD
is ${\cal O} (v^4)$. Additional terms for NREFT Lagrangian in $v^4$
can be systematically studied by including
\begin{widetext}
\begin{eqnarray*}
\label{LNRQCD_2}
\delta {\cal L} = &&\hm - \frac{c_1}{8M^3} \left[\psi^\dagger ({\mathbf D^2})^2
  \psi - \chi^\dagger ({\mathbf D^2})^2 \chi \right] + c_2
\frac{ig}{8M^2}\left[\psi^\dagger \left({\mathbf D}\cdot{\mathbf E} -
  {\mathbf E}\cdot{\mathbf D}\right) \psi + \chi^\dagger \left({\mathbf
    D}\cdot {\mathbf E} - {\mathbf E}\cdot{\mathbf D} \right) \chi \right].
\end{eqnarray*}
\end{widetext}

As for the Monte Carlo data of $SU(3)$ lattice gauge fields which are
used in the calculation, they are generated on $16^3 \times 256$
lattice in ``quenched approximation'' at lattice bare coupling $\beta
= \frac{6}{g^2} = 8.751$ using a single plaquette Wilson
action. Multi-hit Metropolis algorithm together with interleaving
over-relaxation algorithm \cite{Kim:1993gc} is used for the gauge
field update. Each configuration is separated by 1000 Monte Carlo
sweeps. To convert a lattice result into a quantity in a physical
unit, one needs a lattice spacing as a function of the bare coupling
constant. In the scaling limit ($N_f = 0$ due to quenched
approximation),
\begin{equation}
\label{Lambda_L}
a {\Lambda_L}^0 = \exp (- \; \frac{1}{2b_0 g^2}) (b_0
g^2)^{- \; \frac{b_1}{2{b_0}^2}} = f(g) \rightarrow a^{-1} = \frac{{\Lambda_L}^0}{f(g)}
\end{equation}
where $b_0 = \frac{11}{3} \frac{N_c}{16\pi^2}$ and $\frac{34}{3}
(\frac{N_c}{16\pi^2})^2$. We use the light hadron spectrum calculation
in \cite{Kim:1993gc} to set the lattice scale since the experimental
value for $1P-1S$ level splitting used in usual quenched lattice
quarkonium calculations is not available for stoponium. In
\cite{Kim:1993gc}, at $6/g^2 = 6.5$, $m_\rho (m_q \rightarrow 0)a =
0.200(4)$. We obtain $a^{-1} = 3.84(8)$ GeV (${\Lambda_L}^0 = 5.10(1)$
MeV) from the lattice $m_\rho a$ and $m_\rho({\rm physical}) =
768.1(5)$ MeV. This scale setting introduces $\sim 2\%$ systematic
error\footnote{ however, this small error bar should be taken with
  caution. The quenched approximation effect in the hadron spectrum
  can be as large as $20 \%$ and the ambiguity due to the quenched
  approximation in the scale setting is far larger. (see
  e.g. \cite{Gray:2005ur} for the question of the scale setting in
  bottomonium system)}. Thus, the scale for $\beta = 8.751$ is set as
$a^{-1} = 50(1) {\rm GeV} = 0.0039(1) {\rm fm}$.

Under the background of these lattice gauge field configurations,
nonrelativistic stop correlators for ${\cal O} (v^2)$ are calculated
using the evolution equation, 
\begin{eqnarray}
\label{corr_v2}
G (\xv, \tau_0) = &&\hm S(\xv), \nn\\
G (\xv, \tau_i) = &&\hm  \left(1 - \frac{H_0}{2k}\right)^k
U_4^\dagger(\xv,\tau_{i-1}) \left(1 - \frac{H_0}{2k}\right)^k
G(\xv,\tau_{i-1}), \nonumber \\
\end{eqnarray}
where $S(\xv)$ denotes an appropriate complex valued random point
source, diagonal in $SU(3)$ color (random source improves the signal
to noise ratio), and $H_0$ is the lattice Hamiltonian corresponding to
Eq. \eqref{LNRQCD} and $U_4 (\xv,\tau_{i-1})$ is the time directional
gauge field. The parameter, $k$ is introduced to stabilize large
momentum behavior of the lattice discretized evolution equation (see
Table \ref{tab:meanfield} for the choice of $k$). The gauge link
variables are divided by ``tadpole factor'', $u_0$, which is chosen to
be
\begin{equation}
u_0 = \langle 0 \; | \frac{1}{3} {\rm Tr} U_{\rm plaq} | \; 0 \rangle^{\frac{1}{4}}
\end{equation}
where $U_{\rm plaq}$ is a plaquette. For ${\cal O} (v^4)$ Lagrangian,
a modified evolution equation which includes the Hamiltonian for
Eq. \eqref{LNRQCD_2} together with the improvement term for finite
lattice spacing \cite{Lepage:1992tx} can be used such as
\begin{eqnarray}
\label{corr_v4}
G (\xv, \tau_0) = &&\hm S(\xv), \nn\\
G (\xv, \tau_1) = &&\hm  \left(1 - \frac{H_0}{2k}\right)^k
U_4^\dagger(\xv, 0) \left(1 - \frac{H_0}{2k}\right)^k G(\xv,0), \nn \\
G (\xv, \tau_i ) = &&\hm  \left(1 - \frac{H_0}{2k}\right)^k
U_4^\dagger(\xv, \tau) \left(1 - \frac{H_0}{2k}\right)^k \nonumber\\
&&\times \left(1 -\delta H \right)  G (\xv, \tau_{i-1}),
\;\;\;\;\;\;\; (i \ge 2) ,
\end{eqnarray}
where $\delta H$ denotes the lattice $v^4$ Hamiltonian mentioned in the
above.

The zero-momentum stoponium correlators are then formed by combining
the Green function for stop and that for anti stop and are
summed over the spatial lattice sites. From the stoponium correlators,
matrix elements are obtained by fitting
\begin{eqnarray}
\label{corr_fit}
G_{S} (\tau) &=& \sum_n e^{- E_n \tau} |\langle 0 | \chi^\dagger \psi | n
\rangle |^2 \nonumber \\
&=& A_0 e^{-E_{1S} \tau} + A_1 e^{-E_{2S} \tau} + \cdots .
\end{eqnarray}
For the lattice determination of the wave-function at the origin,
$|R(0)|^2$, the matrix element obtained from fitting
Eq. \eqref{corr_fit} is related to the nonrelativistic Coulomb-gauge
fixed wave-function 
\begin{equation}
A_0 = |\langle 0 | \chi^\dagger \psi | 1S \rangle |^2 .
\label{matrixnorm}
\end{equation}
in the leading order of $v^2$ with the normalization convention for
the radial wavefunction, $\int_0^\infty d r \; r^2 \; |R(r)|^2 = 1$
\cite{Thacker:1990bm,Bodwin:1994jh}.

In NREFT formulation, the energy scale in the spectrum is not known.
In order to determine the mass of a state from the energy of state for
a given channel from Eq. \eqref{corr_fit}, ``energy shift'' needs to
be determined,
\begin{equation}
M_N = 2 (Z_M M - E_0) + E_N 
\label{mass_energy}
\end{equation}
where the mass of state, $M_N$ ($N$ denotes the quantum number of the
state), is given in term of the energy of the state ($E_n$), ``zero
point energy ($E_0$)'', and the mass renormalization ($Z_M M$)
\cite{Davies:1994mp} (or one can use the kinetic mass of
nonrelativistic dispersion relation to determine the mass of a state
\cite{Gray:2005ur}). In lattice NRQCD study of charmonium and
bottomonium, this energy shift is determined either by the well
measured experimental mass for one of the states (such as $J/\psi$ and
$\Upsilon$) or by lattice perturbation theory computation
\cite{Lee:2013mla}. Stoponium is not discovered yet and the
perturbative estimate for the energy shift using the lattice action
used in this work is not calculated. Thus, we use tadpole improved
mean field estimate for the energy shift in leading order
\cite{Lepage:1992tx}, where
\begin{equation}
E_0 = - a^{-1} \ln \left[u_0 \left(1 - \frac{a h_0}{2k}\right)^{2k}
  \right], \;\;\; Z_M = u_0^{-1} \left(1 - \frac{a h_0}{2k} \right)
\end{equation}
with
\begin{equation}
a h_0 = 3 \frac{(1 - u_0)}{Ma} .
\end{equation}
$n$ is the integer parameter which is introduced in
Eq. \eqref{corr_v2} to prevent the high momentum instability of the
evolution equation. At $\beta = 8.751 (g^2 = 1.4585)$, Monte Carlo
simulation gives $u_0 = 0.930049(1)$. Perturbatively, $u_0 \sim 1 -
0.083 g^2 - \cdots$ \cite{Lepage:1992tx} and $u_0 \approx 0.94309$ at
$\beta = 8.751$.
\begin{table}[ht]
\begin{center}
\vspace*{0.2cm}
\begin{tabular}{|c|c|c|c|c|c|}
\hline
\hline
$M a$ & $k$ & $ah_0$& $Z_M$ & $E_0 a$ & $2 (Z_M M a - E_0 a)$ \\
\hline
1.0 & 4 & 0.209385 & 1.04700 & 0.285172 & 1.52367 \\
2.0 & 2 & 0.104926 & 1.04700 & 0.178845 & 3.83034 \\
2.5 & 1 & 0.083941 & 1.03008 & 0.158271 & 4.83339 \\
4.0 & 1 & 0.052463 & 1.04700 & 0.125682 & 8.12470 \\
5.0 & 1 & 0.041971 & 1.05265 & 0.114935 & 10.2966 \\
6.0 & 1 & 0.034975 & 1.05641 & 0.107803 & 12.4613 \\
7.0 & 1 & 0.029979 & 1.05910 & 0.102724 & 14.6219 \\
8.0 & 1 & 0.026232 & 1.06111 & 0.098923 & 16.7799 \\
\hline
\hline
\end{tabular}
\vspace*{0.2cm}
\caption{mean field estimates of the zero energy shift}
\label{tab:meanfield}
\end{center}
\end{table}
The difference between the lattice value of $u_0$ and the perturbative
value of $u_0$ is $\sim 1.4 \%$, which suggests order of magnitude for
nonperturbative effect in tadpole factor, $u_0$ at $\beta =
8.751$. Table \ref{tab:meanfield} summarizes the mean field estimates
of the energy shift for various $M a$ and $n$ used for stoponium
correlator calculation. 

There are various sources for systematic errors with nonrelativistic
lattice calculation of stoponium properties. An analysis similar to
that in lattice NRQCD computation of quarkonium \cite{Lepage:1992tx}
can be applied: (1) relativistic correction, (2) finite lattice
spacing, (3) radiative correction, (4) finite lattice volume effect,
and (5) light-quark vacuum polarization. In this work, there are an
additional sources of errors. First is the determination of the zero
energy shift: due to the rest mass of the particle and its
renormalization, NREFT has a undetermined ``zero energy shift'', which
can be fixed by an experimental mass of one of quarkonium states or by
a perturbative calculation. Our use of a mean field theory
approximation \cite{Lepage:1992tx} for the shift introduces a source
of systematic error. Additional error concerns a bare stop mass
used in the lattice calculation. An accurate calculation of the
stoponium masses requires ``tuning of stop mass'' which compares a
stoponium kinetic mass in nonrelativistic dispersion relation for a
given stop mass with an experimental stoponium mass and changes
stop mass until the kinetic mass is equal to the experimental
stoponium mass. Since stoponium is not discovered yet, this tuning
procedure can not be performed and the lattice result in the following
contains a uncertainty from the imprecise tuning of stop mass.

\section{Result}
\label{sec:result}

\begin{figure}[ht]
\centering
 \includegraphics[scale=0.3]{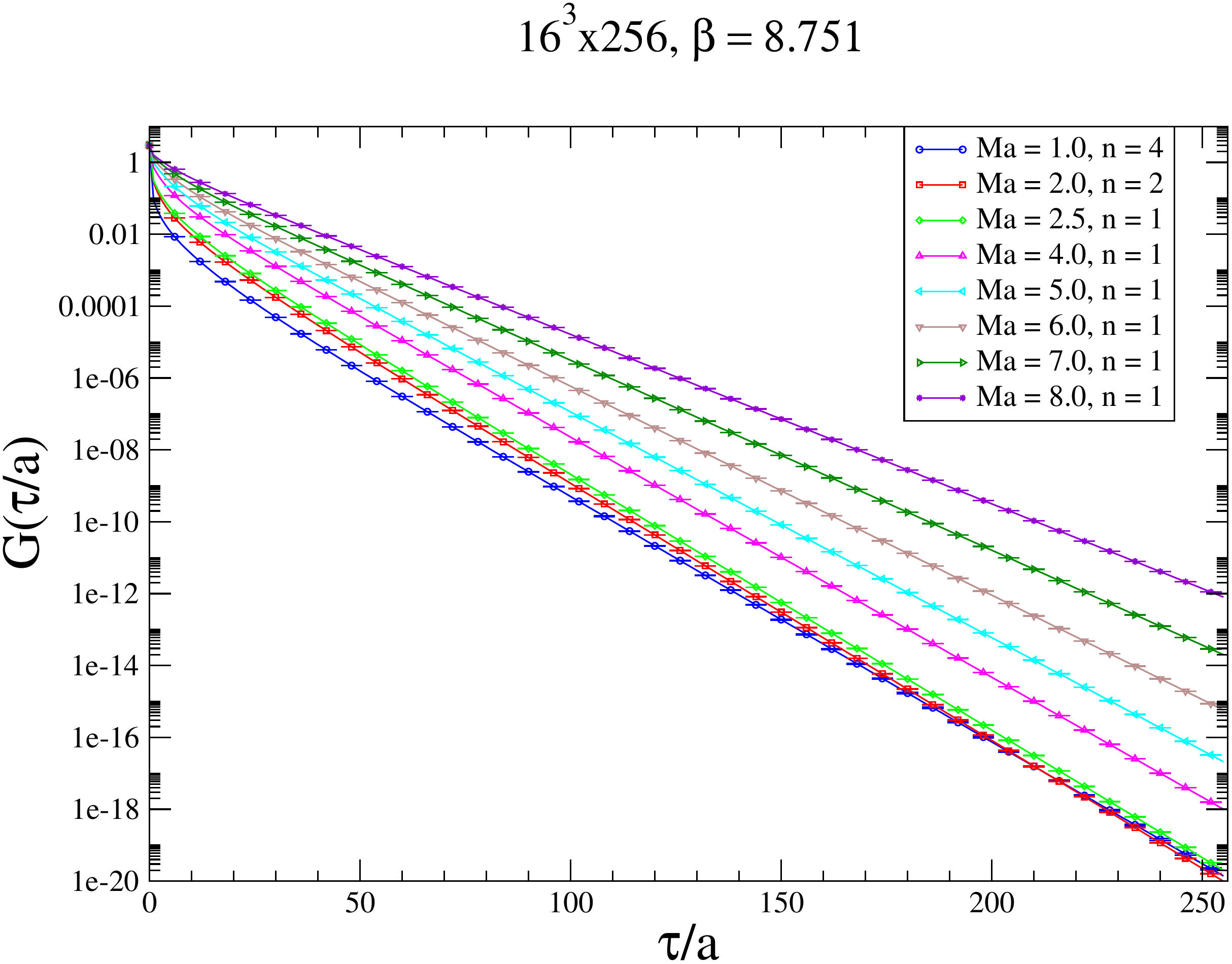}
\caption{The lattice non-relativistic correlators for the S-wave
  channel with ${\cal O} (v^2)$ lagrangian on $16^3 \times 256$ lattices}
 \label{zeroTcorr}
\end{figure}

\begin{figure}[ht]
\centering
 \includegraphics[scale=0.3]{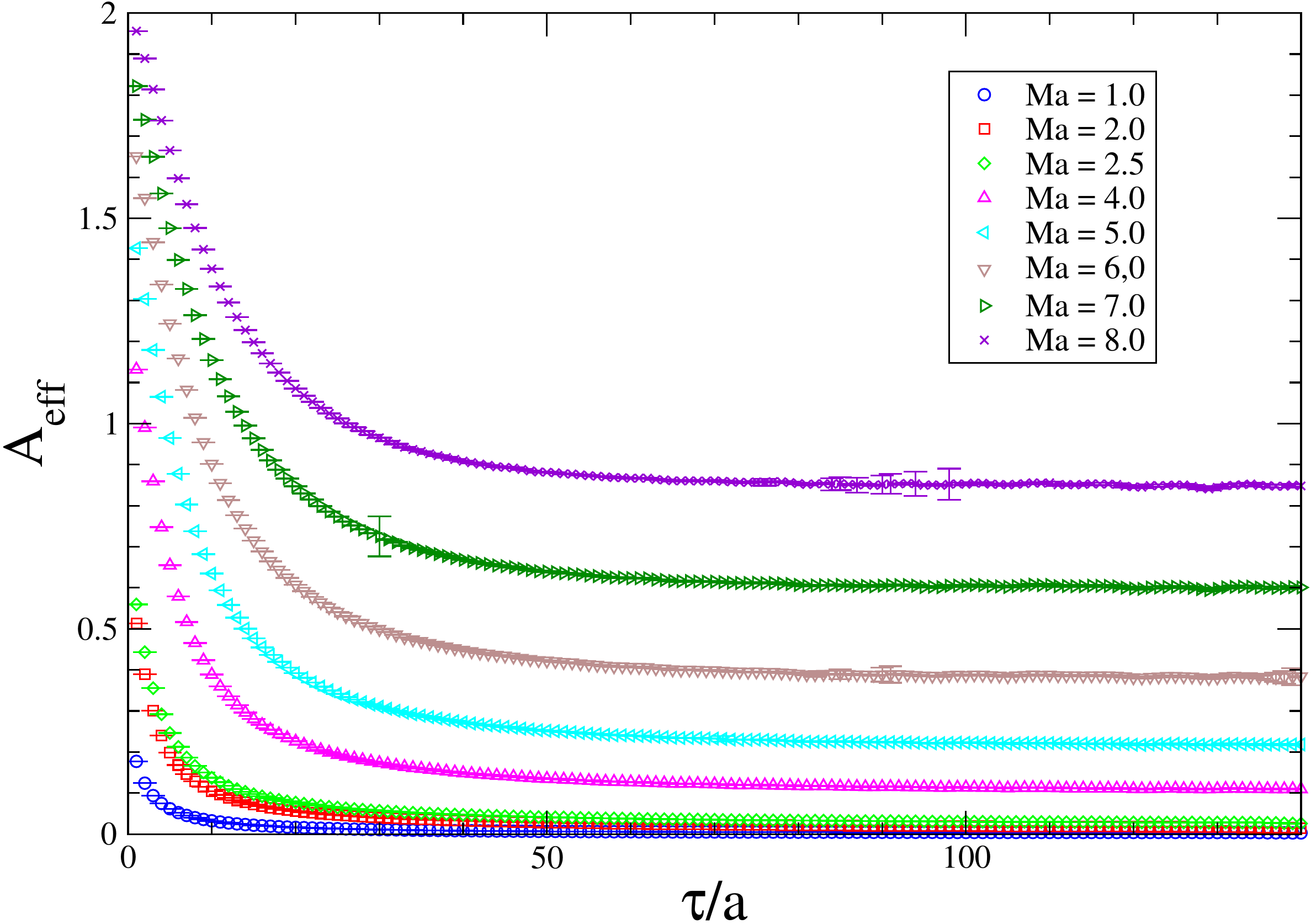}
 \includegraphics[scale=0.3]{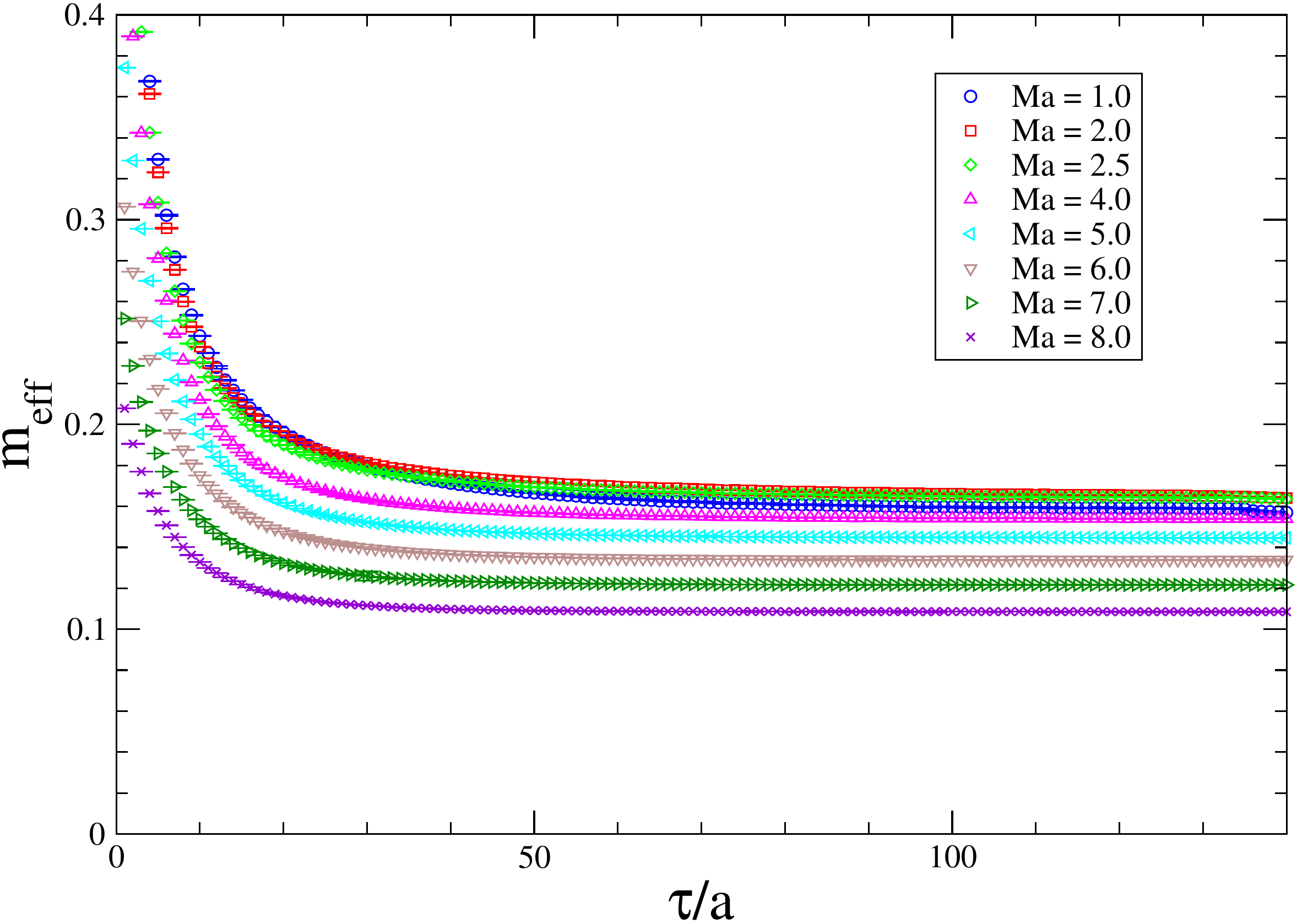}
\caption{The effective mass plot (bottom) and the $A_0$ plot (top) for
  the S-wave channel with ${\cal O} (v^2)$ lagrangian using
  neighboring two points of the correlators ($G(\tau_{i}),
  G(\tau_{i+1})$) on $16^3 \times 256$ lattices. Error bar is from
  Jackknife analysis}
 \label{Ameff}
\end{figure}

Fig. \ref{zeroTcorr} shows S-wave stoponium correlators, $G(\tau)$, on
$16^3 \times 256$ lattice volume calculated with the $v^2$ Lagrangian
(Eq. \eqref{LNRQCD}) and the evolution Eq. \eqref{corr_v2} for each
stop mass $Ma$, where the vertical axis is in the logarithmic
scale. By fitting these stoponium correlators, we obtain the energy of
$1S$ state, $E_{1S} a$ and the amplitude, $A_0 a^3$. Table
\ref{tab:lattice_result_v2} summarizes the fit results for each
stop mass with the fit range $60 \le \tau/a \le 100$ where the fit
range was chosen by locating the plateau region of the effective mass
plot (Fig. \ref{Ameff}). The error bar is from single elimination
Jackknife error analysis of the fitted $E_{1S} a$ and $A_0 a^3$. From
these lattice quantities, $E_{1S} a$ and $A_0 a^3$, we get $M_{1S}$
and $|R_{1S}(0)|^2$ by use of Eq. \eqref{mass_energy} and
Eq. \eqref{matrixnorm} (the last two columns in Table
\ref{tab:lattice_result_v2}).
\begin{table}[ht]
\begin{center}
\vspace*{0.2cm}
\begin{tabular}{|c|c|c|c|c|c|}
\hline
\hline
$M a$ & $k$ & $E_{1S} \; a$& $A_0 \; a^3 $ & $M_{1S}$ (GeV) & $|R(0)|^2/M_{1S}^3$\\
\hline
1.0 & 4 & 0.1619(2) & 0.00507(4) & 84.28(1)  & $4.43(5) \times 10^{-3}$ \\
2.0 & 2 & 0.1688(1) & 0.02377(6) & 200.0(1) & $1.557(7) \times 10^{-3}$ \\
2.5 & 1 & 0.1671(1) & 0.03599(7) & 250.0(1) & $1.205(5) \times 10^{-3}$ \\
4.0 & 1 & 0.1553(1) & 0.1237(2) & 414.0(1) & $9.13(3) \times 10^{-4}$\\
5.0 & 1 & 0.1455(1) & 0.2353(3) & 522.1(1) & $8.66(3) \times 10^{-4}$\\
6.0 & 1 & 0.1344(1) & 0.4014(4) & 629.8(1) & $8.41(3) \times 10^{-4}$\\
7.0 & 1 & 0.1220(1) & 0.6173(5) & 737.2(1) & $8.07(3) \times 10^{-4}$\\
8.0 & 1 & 0.1087(1) & 0.8613(6) & 844.4(1) & $7.49(3) \times 10^{-4}$\\
\hline
\hline
\end{tabular}
\vspace*{0.2cm}
\caption{$E_{1S}$ and $A_0$ from lattice calculation (in lattice unit)
  on $16^3 \times 256$ lattices with $v^2$ NREFT Lagrangian. The
  result is based on $400$ stoponium correlators and the error bar is
  from the jackknife analysis of the 1-exponential fit}
\label{tab:lattice_result_v2}
\end{center}
\end{table}

\begin{figure}[ht]
\centering
 \includegraphics[scale=0.3]{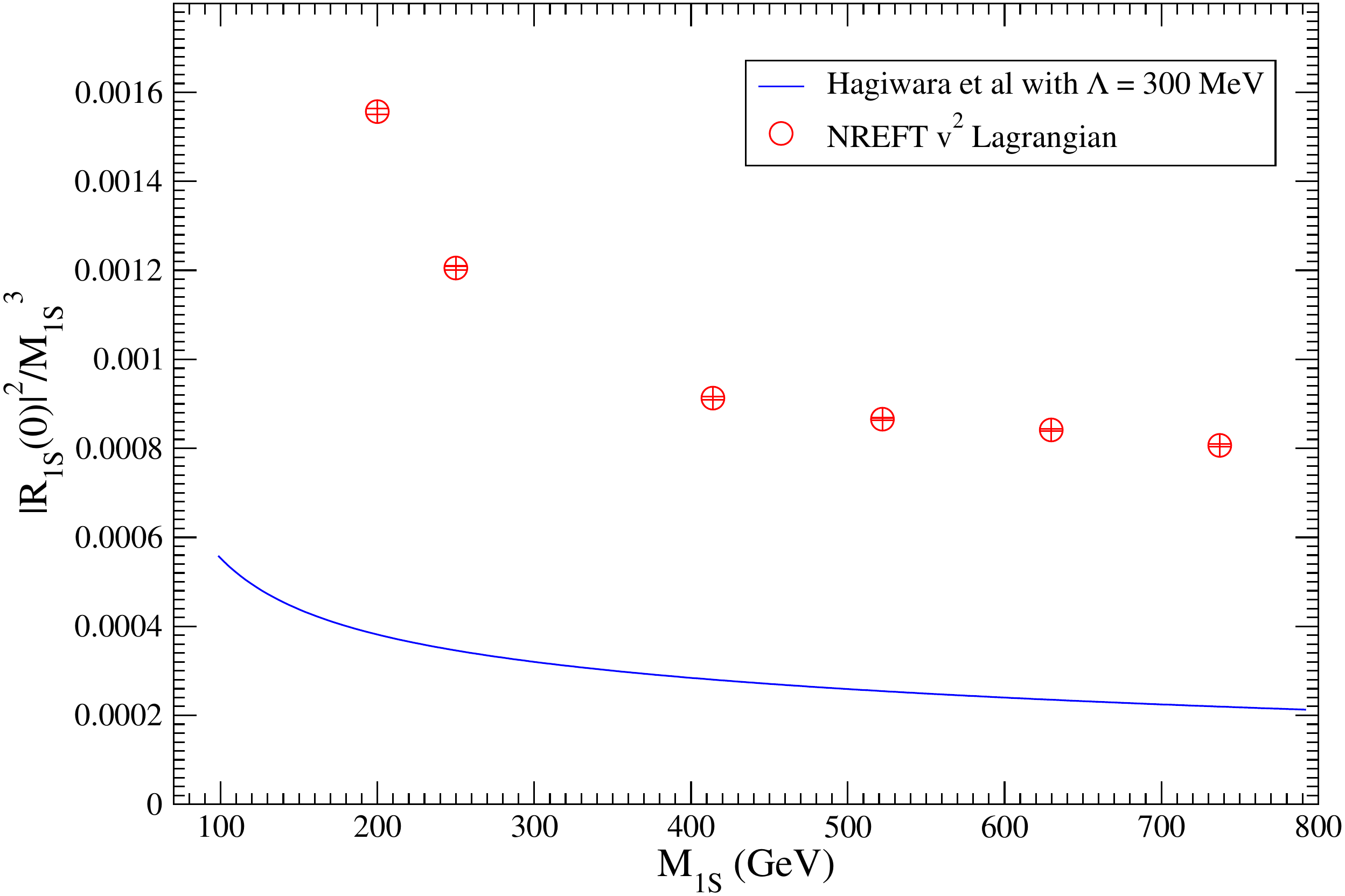}
\caption{$|R_{1S} (0)|^2/M_{1S}^3$ from nonrelativistic correlators
  for the S-wave channel on a $16^3 \times 256$ lattice with ${\cal
    O} (v^2)$ lagrangian} 
\label{matrix}
\end{figure}

Figure \ref{matrix} shows lattice $|R_{1S} (0)|^2/M_{1S}^3$ (the column 6
of Table \ref{tab:lattice_result_v2}) as a function of lattice $M_{1S}$
(the column 5 of Table \ref{tab:lattice_result_v2}) in the range of $100
{\rm GeV} \leq M_{1S} \leq 800 {\rm GeV}$. In the figure, a line for a
potential model result from \cite{Hagiwara:1990sq} is also drawn for a
comparison, where the $\Lambda = 300$ MeV parameterization for $M_{1S}$
and $|R_{1S} (0)|^2/M_{1S}^2$ is used. The figure shows that the
result from lattice NREFT calculation is larger by factor $\sim 4$ at
$M_{1S} \sim 200$ GeV and by $\sim 3.5$ at $M_{1S} \sim 800$ GeV than
that from a potential model calculation.

Let us consider the magnitude of the systematic errors in our lattice
calculation of stoponium. First we consider the finite spacetime
volume effect in Table \ref{tab:lattice_result_v2} by comparing with
the results from two different lattice volumes,
Table \ref{tab:large_lattice_result_v2} from $20^3 \times 256$ lattices
and Table \ref{tab:small_lattice_result_v2} from $12^3 \times 256$
lattices (both at $\beta = 8.751$ with the $v^2$ Lagrangian). These
three tables show that between $16^3 \times 256$ and $20^3 \times
256$, there is little lattice volume dependence in $M_{1S}$ and
$|R_{1S} (0)|^2/M_{1S}^3$ but between $16^3 \times 256$ and $12^3
\times 256$, there are some lattice volume dependences. For $Ma \ge
4.0$, results from two larger lattice volumes agree within error
bar. For $Ma < 4.0$, $M_{1S}$ shows no difference and $|R_{1S}
(0)|^2/M_{1S}^3$ has $\sim 3$ \% difference between the two larger
lattices. Thus, we can conclude that the finite volume effect is
small in the result from $16^3 \times 256$ lattices.

\begin{table}[ht]
\begin{center}
\vspace*{0.2cm}
\begin{tabular}{|c|c|c|c|c|c|}
\hline
\hline
$M a$ & $k$ & $E_{1S} \; a$& $A_0 \; a^3 $ & $M_{1S}$ (GeV) & $|R(0)|^2/M_{1S}^3$\\
\hline
1.0 & 4 & 0.1622(1) & 0.00483(1) & 84.29(1)  & $4.22(1) \times 10^{-3}$ \\
2.0 & 2 & 0.1693(1) & 0.02438(4) & 200.0(1) & $1.596(5) \times 10^{-3}$ \\
2.5 & 1 & 0.1672(1) & 0.03630(6) & 250.1(1) & $1.216(4) \times 10^{-3}$ \\
4.0 & 1 & 0.1553(1) & 0.1233(1) & 414.0(1) & $9.09(2) \times 10^{-4}$\\
5.0 & 1 & 0.1455(1) & 0.2346(1) & 522.1(1) & $8.63(3) \times 10^{-4}$\\
6.0 & 1 & 0.1343(1) & 0.4007(3) & 629.8(1) & $8.40(3) \times 10^{-4}$\\
7.0 & 1 & 0.1220(1) & 0.6171(3) & 737.2(1) & $8.07(2) \times 10^{-4}$\\
8.0 & 1 & 0.1087(1) & 0.8617(4) & 844.4(1) & $7.49(2) \times 10^{-4}$\\
\hline
\hline
\end{tabular}
\vspace*{0.2cm}
\caption{$E_{1S}$ and $A_0$ from lattice calculation (in lattice unit)
  on $20^3 \times 256$ lattices with $v^2$ NREFT Lagrangian. The
  result is based on $400$ stoponium correlators and the error bar is
  from the jackknife analysis of the 1-exponential fit}
\label{tab:large_lattice_result_v2}
\end{center}
\end{table}

\begin{table}[ht]
\begin{center}
\vspace*{0.2cm}
\begin{tabular}{|c|c|c|c|c|c|}
\hline
\hline
$M a$ & $k$ & $E_{1S} \; a$& $A_0 \; a^3 $ & $M_{1S}$ (GeV) & $|R(0)|^2/M_{1S}^3$\\
\hline
1.0 & 4 & 0.1526(4) & 0.0048(4) & 83.81(2)  & $4.27(39) \times 10^{-3}$ \\
2.0 & 2 & 0.1635(2) & 0.0193(2) & 199.7(1) & $1.27(2) \times 10^{-3}$ \\
2.5 & 1 & 0.1638(1) & 0.0310(2) & 249.9(1) & $1.040(9) \times 10^{-3}$ \\
4.0 & 1 & 0.1549(1) & 0.1201(2) & 414.0(1) & $8.86(3) \times 10^{-4}$\\
5.0 & 1 & 0.1455(1) & 0.2328(4) & 522.1(1) & $8.56(3) \times 10^{-4}$\\
6.0 & 1 & 0.1344(1) & 0.3994(6) & 629.8(1) & $8.37(3) \times 10^{-4}$\\
7.0 & 1 & 0.1220(1) & 0.6155(8) & 737.2(1) & $8.04(3) \times 10^{-4}$\\
8.0 & 1 & 0.1087(1) & 0.8592(9) & 844.4(1) & $7.47(3) \times 10^{-4}$\\
\hline
\hline
\end{tabular}
\vspace*{0.2cm}
\caption{$E_{1S}$ and $A_0$ from lattice calculation (in lattice unit)
  on $12^3 \times 256$ lattices with $v^2$ NREFT Lagrangian. The
  result is based on $400$ stoponium correlators and the error bar is
  from the jackknife analysis of the 1-exponential fit}
\label{tab:small_lattice_result_v2}
\end{center}
\end{table}

As discussed, in this work stop mass is not tuned and a mean field
estimate of the zero energy shift is used. These two aspects are
related to each other since tuning amounts to changing lattice stop
mass until the kinetic mass in the nonrelativistic dispersion relation
is equal to the mass of the a given state where part of the ``correct
mass'' is from the zero energy shift, $2(Z_M M - E_0)$. However,
tuning stop mass imprecisely is not a big source of systematic
error. For example, in a bottomonium study \cite{Aarts:2014cda}, the
difference between a properly tuned bottom quark mass, $Ma = 2.92$ and
a rough estimate, $M a = 2.90$ (using $M_b = 4.65$ GeV
\cite{Beringer:1900zz} and the lattice spacing $0.1127$ fm) is
consistent compared with the accuracy of our calculation. From Table
\ref{tab:meanfield}, one observes that the mean field estimate for the
mass renormalization effect is small ($Z_M - 1 \sim 0.05$) at $\beta =
8.751$ and the variation in $E_0 a$ due to change of $Ma$ is $\sim
0.02$ for $Ma \le 4.0$ and $\sim 0.005$ for $Ma > 4.0$. Furthermore
the difference between the leading order perturbative estimate for the
tadpole factor $u_0$ and the Monte Carlo estimate is $1.4 \%$. Thus,
the mean field estimates themselves must be close to the perturbative
estimates for each quantity.

NREFT Lagrangian is an infinite series expansion in $v^2$ which upto
${\cal O} (v^4)$ is given in Eq. \eqref{LNRQCD} and
Eq. \eqref{LNRQCD_2}. Each term other than the kinetic
term\footnote{in this case, tuning of the mass can absorb such
  correction (reparametrization invariance).} (Eq. \eqref{LNRQCD})
comes with effective couplings and radiative corrections to these
coefficients gives series expansions of $\alpha_s$ as $1 + c_i^{(1)}
\alpha_s + {\cal O} (\alpha_s^2)$ for $c_i$ and are expected to be
less than $10 \%$ since $\alpha_s (M) \sim {\cal O} (0.1)$ for $M >
100$ GeV. Similarly, the discretization error due to the finite lattice
spacing must be small since the magnitude of the improvement terms
which correct for the finite lattice spacing effect of the spatial
derivative ($\alpha_s \frac{a^2 \sum_i {p_i}^4}{12 M}/
\frac{p^2}{M}\sim \alpha_s \frac{(Ma)^2 v^2}{12}$) and the temporal
derivative ($\alpha_s \frac{a (\sum_i {p_i}^2)^2}{8 k M^2}/\frac{p^2}{M} \sim
\alpha_s \frac{Ma v^2}{8 k}$) are 
small.

Estimating effects of the quenched approximation on the matrix element
is difficult. In the bottomonium case, the comparison of S-wave
function at the origin from the dynamical simulation
\cite{Bodwin:2001mk} and that from the quenched simulation
\cite{Bodwin:1996tg} using leading order NRQCD lagrangian found that
the matrix element from the quenched approximation underestimates by
$\sim 40 \%$ since the distance scale associated with the bottomonium
bound state ($\sim \frac{1}{M_b v}$) is larger than the scale at which
the matrix elements sample the wave function ($\sim a =
\frac{1}{M_b}$).

In the lattice calculation of the matrix elements, the factorization
scale is related to the lattice cutoff and the effective cutoff is
affected by the specific form of the lattice action and the evolution
equation of lattice Green's functions \cite{Bodwin:2005gg}. Such an
effect needs to be studied if we are interested in the lattice matrix
elements beyond the leading order of nonrelativistic expansion. Also,
since the matrix element in Table \ref{tab:lattice_result_v2} is
actually a lattice matrix element, one needs to calculate perturbative
matching coefficients between the lattice regularization scheme and a
continuum regularization scheme which is used for the parton level
cross section (e.g., $\overline{\rm MS}$) to obtain continuum matrix
elements. In this leading order NREFT Lagrangian study, the matching
is not performed. However, since $\alpha_s$ is small, we expect that
renormalization effect will be small.

\section{Summary}
\label{sec:summary}

Under the assumption that stop is long-lived enough to form bound
states, stoponium plays an important role in the study of stop and
offers an interesting probe to stop searches \cite{Kumar:2014bca,
  Batell:2015zla}. In this case, nonperturbative quantity, $|R_{1S}
(0)|^2/M_{1S}^3$ where $M_{1S}$ is the mass of $1S$ state of stoponium
and $R_{1S} (0)|$ is the radial part of the wavefunction at the
origin, naturally appear in the study of productions and decays of
stoponium. Thus far, a potential model estimate for
$|R_{1S}(0)|^2/M_{1S}^3$ is used in phenomenological investigations of
stoponium, which is unsuitable for improved perturbative calculations.

In this work, using lattice nonrelativistic formulation for heavy stop
which is interacting through the strong interaction, we calculate the
mass of S-wave stoponium and the stoponium matrix element relevant for
stop--anti stop annihilation decay, where this matrix element is
related to $|R_{1S} (0)|^2$ in NREFT scheme. Compared to potential
model studies, lattice study of stoponium is advantageous in that a
particular functional form of the potential between stop and anti stop
needs not be assumed and errors associated with a lattice calculation
can be systematically studied and improved.

Monte Carlo samples of $SU(3)$ color gauge field at $\beta = 8.751$
($a^{-1} = 50 {\rm GeV}$) on $16^3 \times 256$ lattices (which is
generated in quenched approximation with multi-hit
Metropolis/over-relaxation algorithm) is used for lattice
calculation. Table \ref{tab:lattice_result_v2} and Fig. \ref{matrix}
summarizes the lattice result for the $1S$ stoponium mass and the
lattice matrix element for $|R_{1S}(0)|^2/M_{1S}^3$ with ${\cal
  O}(v^2)$ NREFT Lagrangian. In general, the lattice
$|R_{1S}(0)|^2/M_{1S}^3$ is factor $3.5 \sim 4$ larger than a
potential model estimate of \cite{Hagiwara:1990sq} for $200 \leq
M_{1S} \leq 800 {\rm GeV}$. The difference between the lattice result
and the potential model result is larger at lighter $M_{1S}$ and
becomes smaller at heavier $M_{1S}$. According to this trend, the
lattice result may approach Coulombic behavior of the wavefunction at
heavier $M_{1S} > 1 {\rm TeV}$ although upto $\sim 800$ GeV the
lattice result is still far from reaching Coulombic limit. $E_{1S}$,
the energy of $1S$ state of stoponium ranges from $\sim 6$ GeV ($Ma =
1$) to $\sim 8$ GeV ($Ma = 8$). Upto $v^2$ order, toponium is
equivalent to stoponium except the spin degeneracy and the result in
Table \ref{tab:lattice_result_v2} is equally applicable to toponium
since the spin-flip term in NRQCD is ${\cal O} (v^4)$
\cite{Bodwin:1994jh}.

Although further studies on systematic effects in our lattice result
is necessary, factor $3.5 \sim 4$ larger $|R_{1S}(0)|^2/M_{1S}^3$
implies an enhanced stoponium production cross section in hadron
colliders such as LHC and can have interesting consequences for stop
search in hadron colliders. Therefore, we hope to study stoponium
using $SU(3)$ lattice gauge fields which includes vacuum polarization
effect of light dynamical quark and improve the above result using
${\cal O} (v^4)$ NREFT Lagrangian in the future. For a next-to-leading
order NREFT Lagrangian study, other matrix elements which are higher
order in $v^2$ is also needed and the perturbative matching mentioned
in the above needs to be performed. Since excited states of S-wave
stoponium will contribute to the production cross section, matrix
elements for excited states will be also interesting to study.

\FloatBarrier
\section*{Acknowledgments}
We would like to thank C.T.H. Davies, R.R. Horgan for discussion
during Royal Society International Scientific Seminars at Chicheley
Hall, ``Heavy quarks: a continuing probe of the strong
interaction''. Discussions with Sunghoon Jung, Pyungwon Ko on
stoponium and Chaehyun Yu, Jungil Lee on quarkonium were helpful. This
work is supported by the National Research Foundation of Korea grant
funded by the Korean government (MEST) No.\ 2015R1A2A2A01005916 and in
part by NRF-2008-000458.

\end{document}